# Pressure-induced superconductivity in $La_4Ni_3O_{10+\delta}$ ($\delta$ = 0.04 and −0.01)

Hibiki Nagata[1,2*], Hiroya Sakurai[1], Yuta Ueki[1,2], Kazuki Yamane[1,2], Ryo Matsumoto[1], Kensei Terashima[1], Keisuke Hirose[3‡], Hiroto Ohta[3], Masaki Kato[3], and Yoshihiko Takano[1,2†]

[1]*MANA, National Institute for Materials Science, 1-2-1 Sengen, Tsukuba, Ibaraki 305-0047, Japan*

[2]*Graduate School of Pure and Applied Sciences, University of Tsukuba, 1-1-1 Tennodai, Tsukuba, Ibaraki 305-8577, Japan*

[3]*Faculty of Science and Engineering, Doshisha University, 1-3 Tataramiyakotani, Kyo-Tanabe 610-0321, Japan*

The superconducting transition temperatures, $T_c$, of $La_4Ni_3O_{10+\delta}$ ($\delta$ = 0.04 and −0.01) were determined under various pressures up to 124.9 GPa by electrical resistance measurements with a diamond anvil cell. $T_c$ exhibits a strong dependence on oxygen content within the pressure range of approximately 20 GPa and 80 GPa. At 48.0 GPa, $T_c$ of $La_4Ni_3O_{10.04}$ peaks at 36 K, marking the highest $T_c$ reported thus far.

The recent discovery of superconductivity in $La_3Ni_2O_7$ has made a significant impact, not only because of its high transition temperature ($T_c$ ~ 80 K under 14 GPa)[1] but also due to its structural resemblance to high-$T_c$ cuprates. These characteristics have sparked intense research into superconductivity on $NiO_2$ layers from both theoretical and experimental perspectives. A major experimental endeavor has been made to discover new superconductors related to the compound, akin to the approach taken with the high-$T_c$ cuprates, as such discoveries are expected to contribute to the characterization and understanding of superconductivity. In this context, we made the initial discovery that $La_4Ni_3O_{9.99}$, featuring triple $NiO_2$ layers, also undergoes a superconducting transition at 23 K under a high pressure of 79.2 GPa.[2] Although the superconductivity was promptly confirmed by other research groups,[3-6] the reported transition temperatures have shown inconsistency. This has motivated us to investigate a variety of samples with different chemical compositions. As a first trial, we synthesized $La_4Ni_3O_{10.04}$ and measured its electrical resistance under various pressures to compare it with that of $La_4Ni_3O_{9.99}$ additionally measured at higher pressures than in the previous study.[2]

A powder sample of $La_4Ni_3O_{10.04}$ was synthesized using a similar method to that employed for La$_4$Ni$_3$O$_{9.99}$.[2] The only difference lies in the heat treatment during hot isostatic pressing: for the former, the sample was annealed for 1hour at 1200°C under a 400 atm oxygen partial pressure, whereas for the latter, it was annealed for 2hours at 600°C under a 300 atm oxygen partial pressure. The determination of oxygen content was carried out as described in the previous paper.[2]

The samples were characterized by powder X-ray diffraction measurements using a commercial diffractometer, MiniFlex600 (Rigaku). All observed peaks were indexed assuming an orthorhombic crystal structure of space group *Cmca*,[7] as depicted in Fig. 1, indicating that each sample was of a single phase. The lattice parameters were estimated from the peak positions to be $a$ = 5.413 Å, $b$ = 5.465 Å, and $c$ = 27.97 Å for La$_4$Ni$_3$O$_{10.04}$, and $a$ = 5.412 Å, $b$ = 5.464 Å, $c$ = 27.98 Å for La$_4$Ni$_3$O$_{9.99}$.

The electrical resistance of the sample was measured under various pressures up to 124.9 GPa using conventional four-probe method employing a diamond anvil cell (DAC) with boron-doped diamond electrodes.[2, 8-9] Stainless steel as the gasket and cubic BN as the pressure transmitting medium (PTM) were used in the setup of DAC. For the measurements of both samples, the same configuration of DAC and procedures to generate pressure were adopted.

The temperature dependence of the resistance of the samples is shown in Fig. 2. The resistance decreases with increasing pressure for both samples. One may notice that the resistance increases with decreasing temperature under certain conditions. However, we maintain that the samples are metallic at any temperature under the pressures, as evidenced by the absence of divergent behavior toward 0 K. The observed upturns are likely attributed to extrinsic factors, such as grain-boundary scattering, as sometimes seen in the resistance obtained using polycrystalline samples.[3, 10-14]

In the case of La$_4$Ni$_3$O$_{10.04}$, the resistance exhibits a slight suppression below 5 K at 20.2 GPa, with this suppression becoming more pronounced under higher pressures, as depicted in Fig. 2a. The suppression is most likely attributable to the onset of superconductivity because, as shown in Fig. 3, it displayed a similar field dependence to that observed with La$_4$Ni$_3$O$_{9.99}$.[2] For La$_4$Ni$_3$O$_{9.99}$, it is evident that the drop in resistance, which has been attributed to superconductivity below 79.2 GPa,[2] persists even at pressures of 112.6 GPa and 124.9 GPa. Notably, this marks the first observation of superconductivity in La$_4$Ni$_3$O$_{10}$ at pressures exceeding 100 GPa.

The superconducting transition temperature was estimated as the temperature at which the resistance began to deviate from an almost linear temperature dependence just above it, as indicated by the arrows in Fig. 2. These temperatures are summarized in Fig. 4. For La$_4$Ni$_3$O$_{10.04}$,

$T_c$ rapidly increases with increasing pressure from 20.2 GPa, reaches at 36 K at 48.0 GPa, and then gradually decreases with further pressure increase. This $T_c$ of 36 K represents the highest value reported thus far. In contrast, for $La_4Ni_3O_{9.99}$, superconductivity begins to appear at a higher pressure of 32.8 GPa and gradually increases with increasing pressure. Thus, $T_c$ of $La_4Ni_3O_{10}$ exhibits a strong dependence on oxygen-content within the pressure range of approximately 20 GPa and 80 GPa. Because the setup of DAC for both compounds are the same, oxygen content plays a crucial role in the emergence of superconductivity. Interestingly, most reported $T_c$ values thus far fall between those of $La_4Ni_3O_{10.04}$ and $La_4Ni_3O_{9.99}$ within the pressure range of 20 GPa and 80 GPa. [3-6] Although some of the samples used in the previous reports[3-6] were synthesized under lower oxygen partial pressures than $La_4Ni_3O_{9.99}$, the variation in $T_c$ may be attributed to difference in oxygen content. Here, our PTM is cBN which generally provides non-hydrostatic pressure. As a future attempt, the investigation on influence of difference in PTM is needed to reveal more intrinsic effect of oxygen content on superconductivity because sharp superconducting transition can be obtained by applying hydrostatic pressure. [5,15-17]

**Acknowledgment**

We would like to express our gratitude to Professors K. Kuroki (Osaka Univ.), H. Sakakibara (Tottori Univ.), and M. Ochi (Osaka Univ.) for fruitful discussion. This work is supported by JSPS KAKENHI Grants No. JP20H05644 and JP24K01333.

*E-mail: NAGATA.Hibiki@nims.go.jp
†E-mail: TAKANO.Yoshihiko@nims.go.jp
‡On leave from

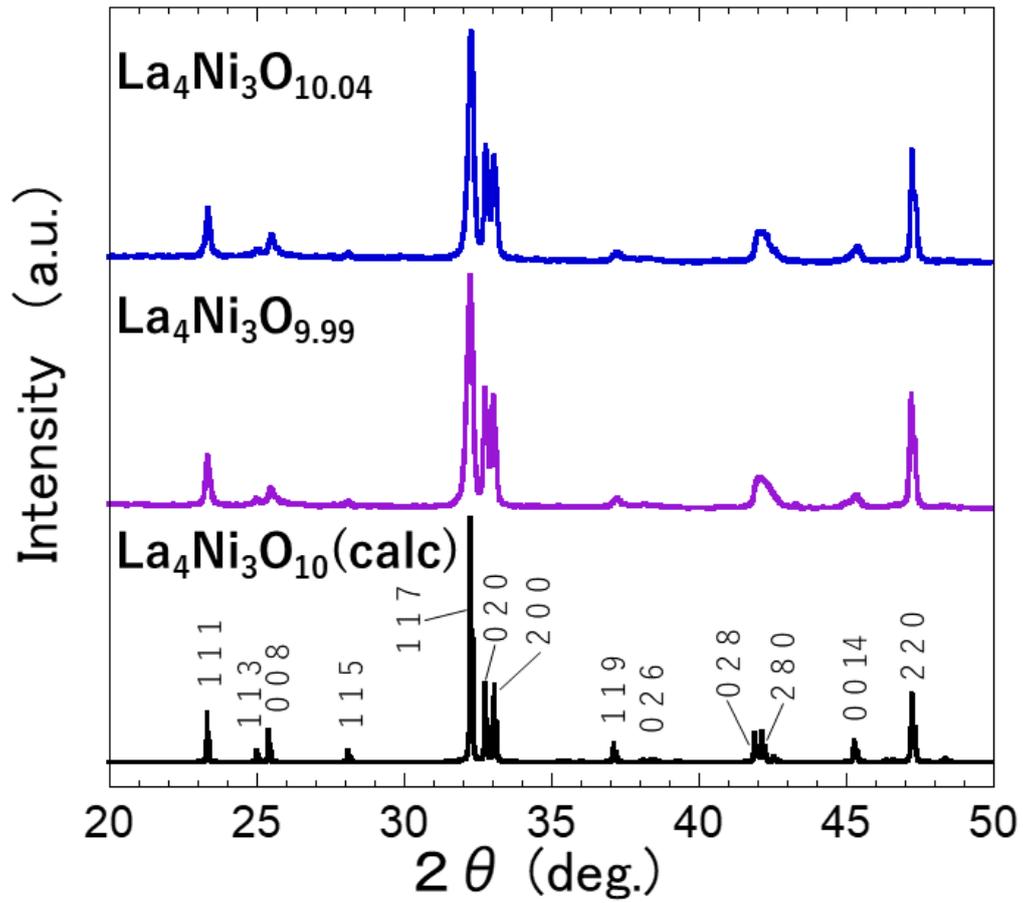

Fig. 1 (Color online) Powder X-ray diffraction patterns of $La_4Ni_3O_{10.04}$ and $La_4Ni_3O_{9.99}$, alongside a simulated pattern from the structural data.[7] Miller indices are shown for some characteristic peaks.

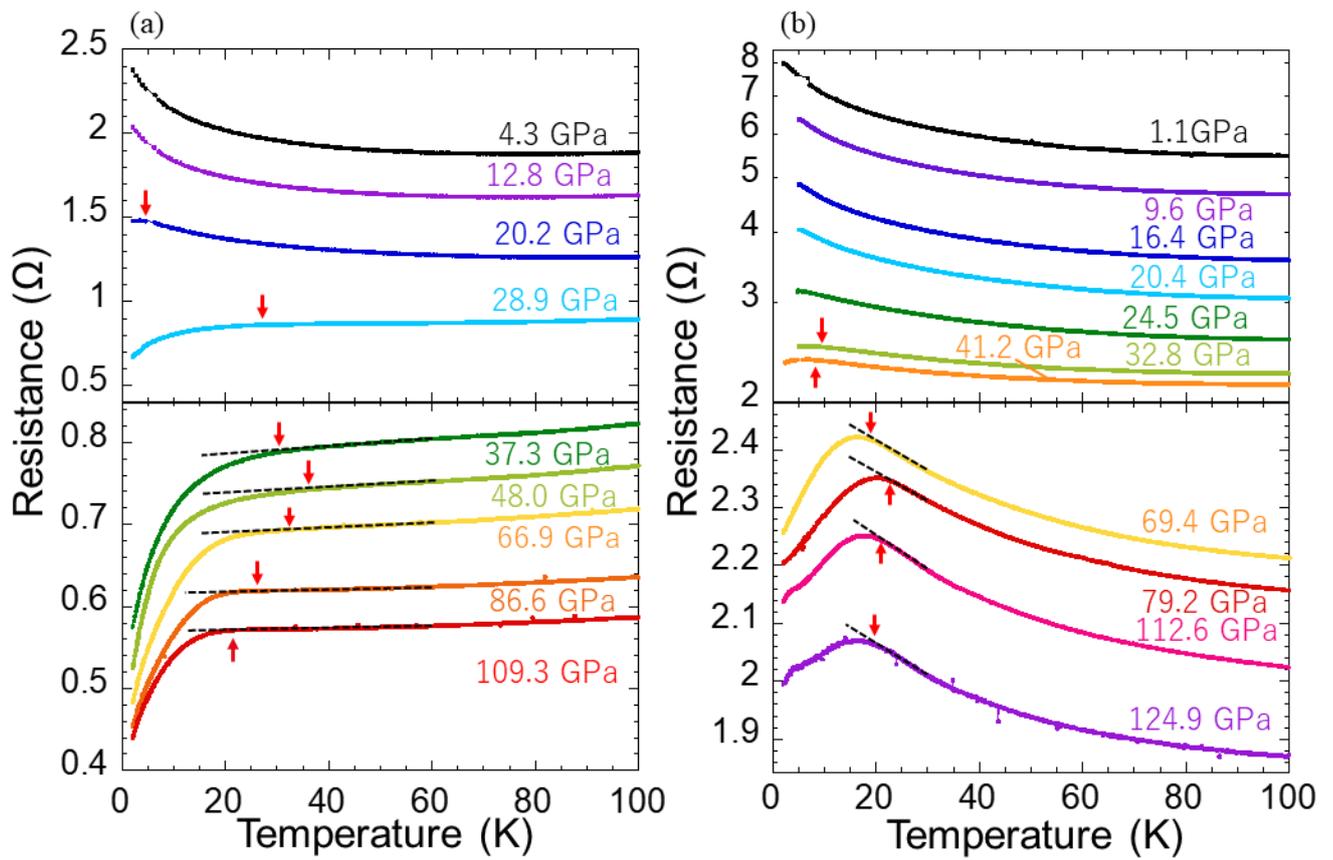

Fig. 2 (Color online) Temperature dependence of the resistance of $La_4Ni_3O_{10.04}$ (*a*) and $La_4Ni_3O_{9.99}$ (*b*) under various pressures. The dashed lines represent the linear temperature dependence just above $T_c$, and the arrows indicate $T_c$ estimated (see the text).

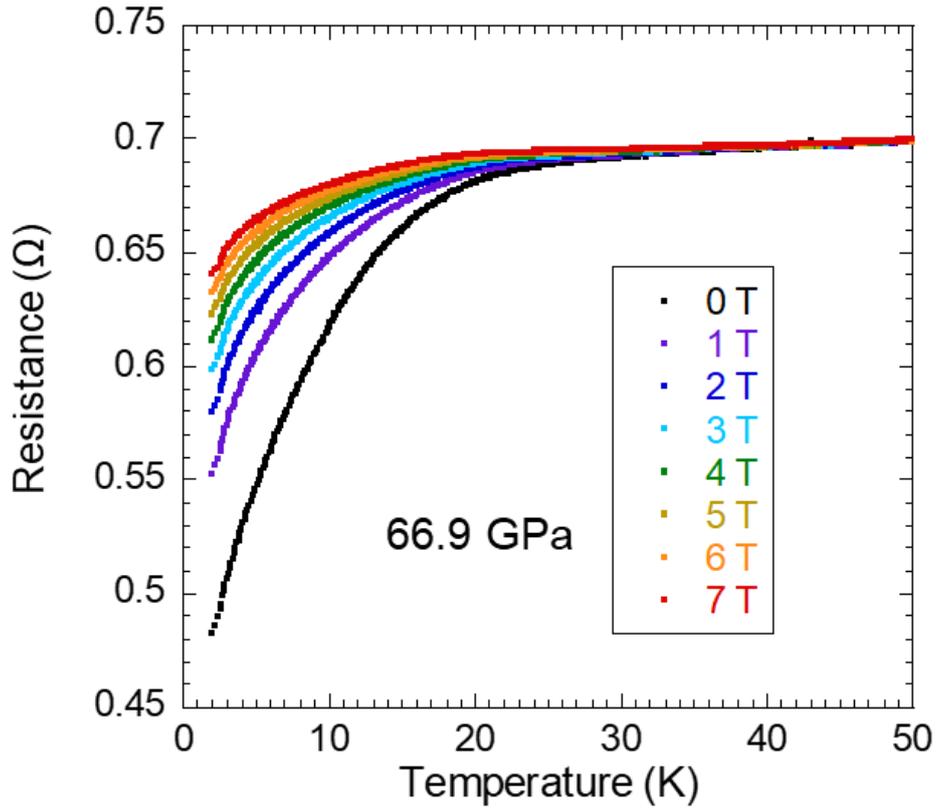

Fig. 3 (Color online) Temperature dependence of the resistance of $La_4Ni_3O_{10.04}$ under magnetic fields between 0 and 7 T under 66.9 GPa.

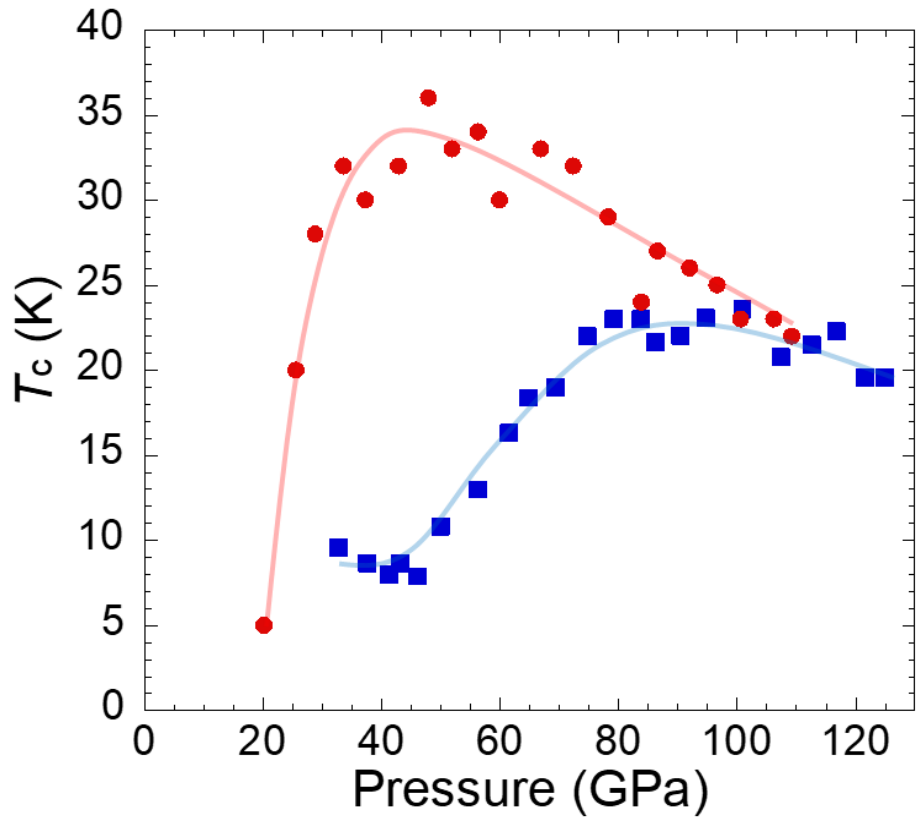

Fig. 4 (Color online) $T_c$ of $La_4Ni_3O_{10.04}$ (red filled circles), $La_4Ni_3O_{9.99}$ (blue filled squares).

The lines are guides to eye.